\documentclass[letterpaper]{article}
\usepackage{nips11submit_e,times}

\usepackage{url}
\usepackage{bm}

\usepackage{multicol}
\usepackage{marginnote}
\usepackage[utf8x]{inputenc}
\usepackage{lmodern}
\usepackage{graphicx}
\usepackage{algorithmic}
\usepackage{algorithm}
\usepackage{amssymb,amsmath}
\usepackage[T1]{fontenc}
\usepackage{xcolor}

\makeatletter
\def\blfootnote{\gdef\@thefnmark{}\@footnotetext}
\makeatother

\newcommand{\namea}{\textsc{Adams}}
\newcommand{\nameb}{\textsc{Jefferson}}
\newcommand{\namec}{\textsc{Madison}}
\usepackage{caption}
\usepackage{subcaption}

\def\ngp#1#2{
  $\vcenter{\hbox{
  \valign{\hbox{\strut##}&
          \hbox{\strut##}\cr
          #1&#2\cr}}}$}
          
\newcommand{\ignore}[1]{}
\nipsfinalcopy 

 \title{Inferring Social Rank \\ in an Old Assyrian Trade Network}

\author{
David Bamman\\
School of Computer Science\\
Carnegie Mellon University\\
\texttt{dbamman@cs.cmu.edu} \\
\And
Adam Anderson\\
Department of Near Eastern\\
Languages and Civilizations\\
Harvard University\\
\texttt{aganders@fas.harvard.edu}
\And
Noah A.~Smith\\
School of Computer Science\\
Carnegie Mellon University\\
\texttt{nasmith@cs.cmu.edu} \\
}

\begin{document}
\maketitle

\section{Introduction}

\blfootnote{\hspace{-16pt}Appearing in \emph{Digital Humanities 2013}, Lincoln, Nebraska.}

In the early 20th century, the attention of Assyriologists and
archaeologists was directed to a number of cuneiform tablets
coming from a remote archaeological tell in Kültepe,
Turkey. After the first series of excavations, archaeologists discovered a large collection of texts and the remains of a Bronze Age trade colony, referred to in the texts as kārum Kaneš. Once these initial ca.~5,000 texts 
were deciphered, the field of Old Assyrian studies was born. In 1948 official Turkish excavations began at Kültepe and added over 17,000 tablets to the Old Assyrian text corpus, which now totals ca.~23,000 cuneiform tablets \cite{michel_Alahum_2008}. These texts document the intricacies of thriving Bronze Age trade networks, comprised of Old Assyrian merchants from the ancient city of Assur approximately 4,000 years ago
 (ca.~1950-1750 BCE) \cite{barjamovic_Ups_2012}. The texts further show how the merchants acted as the middle-men in a large series of inter-connected networks which, among other things, linked the natural resources of tin (in Iran and Afghanistan) and copper (in Turkey) in order to produce bronze in Anatolia.

However, one thing the texts do not make clear is the scope and structure of the colonial trade network, in terms of the people involved and their organization. Although the high degree of literacy among the inhabitants of the colony at Kaneš helped create an extremely rich source of texts illustrating the daily life of the people involved, the practice of paponomy (naming a son after his grandfather) has obscured the identities of the merchants for modern scholarship. Thus, due to the density and ambiguity of the names mentioned in these texts, it has been too difficult to gain an understanding of the scope of the colonial society on the basis of the textual record at Kültepe.

Our work therefore focuses on jointly inferring the unique individuals as well as their social rank within the Old Assyrian trade network, using a novel probabilistic latent-variable model that exploits partial rank information contained in the texts.

\section{Data} \label{se:data}

Of the 23,000 tablets unearthed at Kültepe (Kaneš), 5,691 published texts (known as the ``old texts''\ \cite{veenhof}) have been digitized and transcribed into machine-actionable text as part of the Old Assyrian Text Project\ \cite{oatp}.  While these tablets include mostly economic and legal transactions, 2,094 of them are letters between merchants. Along with the body of the Akkadian text, each of these letters includes a highly stylized introductory formula (an ``epistolary formula'') which lists the senders and recipients using strict dominance rules concerning the order of the names. For example: 

\begin{center}
\ngp{umma}{from}
\ngp{Aššur-idī}{Aššur-idī}
\ngp{Aššur-nādā}{and Aššur-nādā}
\ngp{ana}{to}
\ngp{Amur-Ištar}{Amur-Ištar,}
\ngp{Alāhum}{Alāhum}
\ngp{Aššur-taklāku}{and Aššur-taklāku}
\end{center}

These formulae have a consistent internal structure from which we can draw relative social ranks among the individuals involved.  Each formula can be divided into two parts (a \emph{receiving} rank and a \emph{sending} rank), and an individual placed linearly after another \emph{within} one of these ranks cannot be socially higher than any mentioned before (whether the first is higher or equal is ambiguous).  Additionally, one individual (who may be either among the senders or recipients) is mentioned first in the letter, a marked position signifying the highest social status of those mentioned in either rank.  

\begin{align}
\label{rank}
 \underbrace{\overbrace{\textrm{Aššur-idī}}^{\textrm{first mentioned}}\textrm{+ Aššur-nādā}}_{\textrm{sending rank}} \rightarrow \underbrace{\textrm{Amur-Ištar + Alāhum + Aššur-taklāku}}_{\textrm{receiving rank}}
\end{align}

These partial orderings provide a rich source of evidence for the global social structure; from this example, we can extract seven pairwise partial social orderings: Amur-Ištar  $\ge$ Alāhum and Aššur-taklāku; Alāhum $\ge$ Aššur-taklāku; and Aššur-idī $\ge$ all four of the others.  

If all such partial orderings were to be trusted, if each observed
name in such a formula were unambiguous, and if social power were a
stationary quality that remained constant over time, inferring a
consistent global rank over all individuals would be easy (though more than one such rank may be possible).
Unfortunately, however, none of these assumptions are true.  The rank
we observe in one letter is a subjective judgment by the author, and
we can easily imagine that complex social dynamics are involved in the
choice of who to rank highest (which can vary by author); names are
indeed ambiguous with one name potentially referring to multiple
people, and the same person can be known by several names; and the
letters span a period of ca.~200 years, over which time a young
individual with low social rank can age and accrue
power.\footnote{Most of the letters in our corpus have not been dated to a finer level of granularity than century; a potential extension to the model described here would exploit this information when available.}

\section{Technical Approach}

Our goal is to find the social ranking
over individuals (possibly not in a one-to-one relation with names)
that best explains the observed data. To illustrate the intuition
behind our approach, consider a simple example.  Suppose we are trying
to establish the temporal rank of a set of individuals with the names
{\namea}, {\nameb}  and {\namec}, and we have the following evidence (where $>$ indicates ``was president before'').

\begin{itemize}
\item {\namea} > {\nameb}
\item {\nameb} > {\namec}
\end{itemize}

Assuming transitivity, a sound global rank among these three is: {\namea} > {\nameb} > {\namec}; while we never directly witness a statement of the sort {{\namea} > {\namec}}, we can infer it through intermediary relations.  Now suppose we observe an additional piece of evidence:

\begin{itemize}
\item {\namec} > {\namea}
\end{itemize}

If we assume that the three names only refer to three distinct
individuals, transitivity breaks down: putting all three statements
together results in a circular rank, leading to the contradiction that
{\namea} > {\nameb} while at the same time {\nameb} > {\namea}. However,
we can establish a sound global order if we allow the three names to
refer to four individuals (e.g., two people both have the name
{\namea}), resulting in the rank: \namea$_1$ > {\nameb} > {\namec} >
\namea$_2$.\footnote{Alternatively, \nameb$_1$ > {\namea} > {\namec} > \nameb$_2$ is
  also valid, as is \namec$_1$ > {\nameb} > {\namea} > \namec$_2$. In data where the orderings are not strict (i.e., {\namea} $\ge$ {\nameb}), global ranks involving equalities are also possible.} In fact, given
the inconsistency of the evidence under the assumption of only three
people, the existence of four underlying people is in fact more
likely.   Here, our method offers an informed hypothesis---that {\namea} refers to two distinct individuals rather than one---that can be verified (or refuted) in consultation with the data.  In this simple case, the hypothesis is supported by the fact that \textsc{Adams} can refer to both \textsc{John Adams} and \textsc{John Quincy Adams}.

In the case of our Old Assyrian dataset, the evidence takes the form
of 4,191 pairwise observed ranks of 717 individual names in 1,657
letters, along with the Akkadian text of those letters.  Our task is
to find the most likely overall social rank of a fixed set of actors that best
explains the pairwise ranks in letters that we see.  In casting the problem in this way, we are
building a \emph{probabilistic generative model} of the data.  Latent
Dirichlet allocation~\cite{Blei:2003:LDA:944919.944937} is a familiar
example of a generative model that seeks to explain data in text
documents
by 
inferring latent topic assignments to individual words.  In our case,
the latent variables are a.) the identities of the people named in
each letter; and b.) the social rank of those people, represented as 
continuous values.

Figure 1 illustrates the graphical model in detail using standard
notation alongside our inference algorithm.  In brief, we use a
randomized algorithm known as Monte Carlo Expectation Maximization
\cite{wei90}.  The algorithm alternately a.) samples a value of the
latent entity $z$
for each instance of a name $x$ in a given letter, conditioned on
current values of all other latent variables and parameters, cycling
through the name instances, and b.)
uses those accumulated
samples to optimize the values of the social rank $\beta$ that
generated the pairwise ranks $y$ in evidence.
In this way, we alternate between picking probable latent
individuals referred to in letters (given some fixed social rank), and determining the best social rank given that estimate of who those names refer to.

\begin{figure}[htp]
\begin{centering}
\begin{subfigure}{2.5in}
\includegraphics[scale=1]{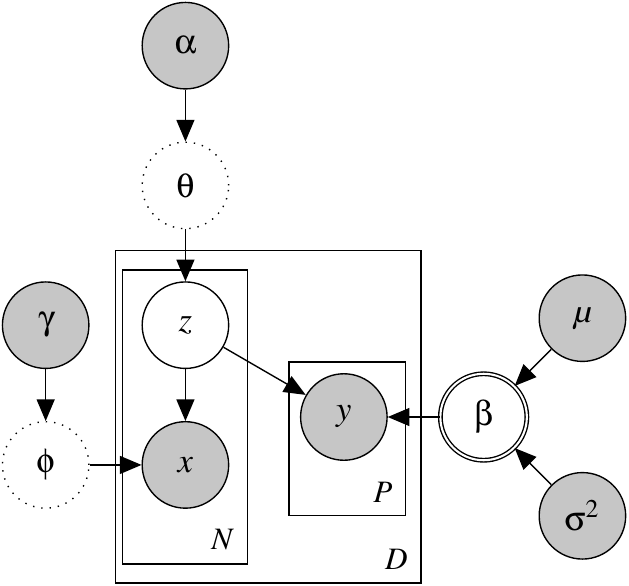}
\end{subfigure}
\quad\quad
\begin{subfigure}{3.5in}
\begin{algorithmic}
\FOR{$i = 1 \to $ numIterations} 
	\STATE samples $\leftarrow \emptyset$
	\FOR{$j = 1 \to $ numSamples} 
		\FORALL {letters $l$ in corpus} 
			\FORALL {names $n$ in $l$} 
				\STATE sample latent entity $z_{n,l} \propto p(\theta|\alpha) p(\phi|\gamma) \times$
				\STATE $ p(\beta|\mu, \sigma^2)\prod_D  p(y|z, \beta)\prod_N p(z|\theta) p(x|z, \phi)$
				\IF{$j > 100$ and $j$ mod $10 = 0$} 
				\STATE samples $\leftarrow$ samples + $z_{n,l}$
				 \ENDIF
			\ENDFOR
		\ENDFOR
	\ENDFOR

	\STATE Maximize $\sum_{\langle z_i, z_j \in \textrm{samples}\rangle} \left(
	  \frac{1}{1+\exp(-(\beta_{z_i} - \beta_{z_j}))} \right) {-\frac{1}{\sigma^2} \| \beta - \mu\|_2^2} $ 
\ENDFOR
 \end{algorithmic}
\end{subfigure}
\caption[]{Graphical model and MCEM algorithm. Shaded circles represent observed or fixed variables, empty circles are latent variables, dotted circles represent variables that are integrated out via collapsed Gibbs sampling, and the double-circled $\beta$ is optimized via MCEM.  $\phi, \theta \sim \textrm{Dir}$;  $z \sim \textrm{Categorical}$;  $x \sim \textrm{Multinomial}$;  $y \sim \textrm{Logistic}$;  and $\beta \sim \textrm{Normal}.$  In our case, we fix the hyperparameters $\alpha=100, \gamma=0.01, \mu=0, \sigma^2 =1$, and the number of possible latent entities to 1000.}
\label{doc}
\end{centering}
\end{figure}

\section{Results}

The input to our algorithm is a set of pairwise ranks between names mentioned in a letter (of the form Aššur-idī $\ge$ Aššur-nādā, as above), along with an upper bound on the number of the latent entities we expect ($K$); the output is twofold: a.) a global rank of those $K$ latent entities, along with the names in letters associated with them most often; and b.) a distribution over all possible latent entities for each name mentioned in each letter.

We apply our model to \emph{hypothesis generation}: given a set of
evidence, the algorithm offers hypotheses it finds likely, which a domain expert can then validate according to established methods in the field.    One such lead generated by our method about a well-studied individual concerns the name of Innāya.
	
\subsection*{Innāya}

In 1991, Cécile Michel produced a two-volume work on two merchants in the colony of Kaneš named Innāya \cite{michel_Innaya_1991a, michel_Innaya_1991b}. On the basis of two attested patronyms it was apparent that there were at least two individuals who were known by this name. 
By charting their family trees and the structure of their respective businesses, she reconstructed the separate archives and identities of these two merchants. The first and best attested individual is Innāya son of Elāli with 142 texts, who appears to have a more complete textual record. The second individual, Innāya son of Amurāya, is only attested in 74 texts; while these two merchants overlapped chronologically, the latter appears to have been a minor figure in the colony \cite[p. 48]{michel_Innaya_1991a}. While there were a number of texts which Michel was unable to determine (ca. 57), her study illustrated the complexity involved in the Old Assyrian archives due to an active use of homonyms at that time. While the work on Innāya is in no way complete, Michel provided a basis on which future scholarship might build, and will serve as a proving ground for the purpose of this study.

\begin{figure}[h*]
\begin{centering}
\includegraphics[scale=.45]{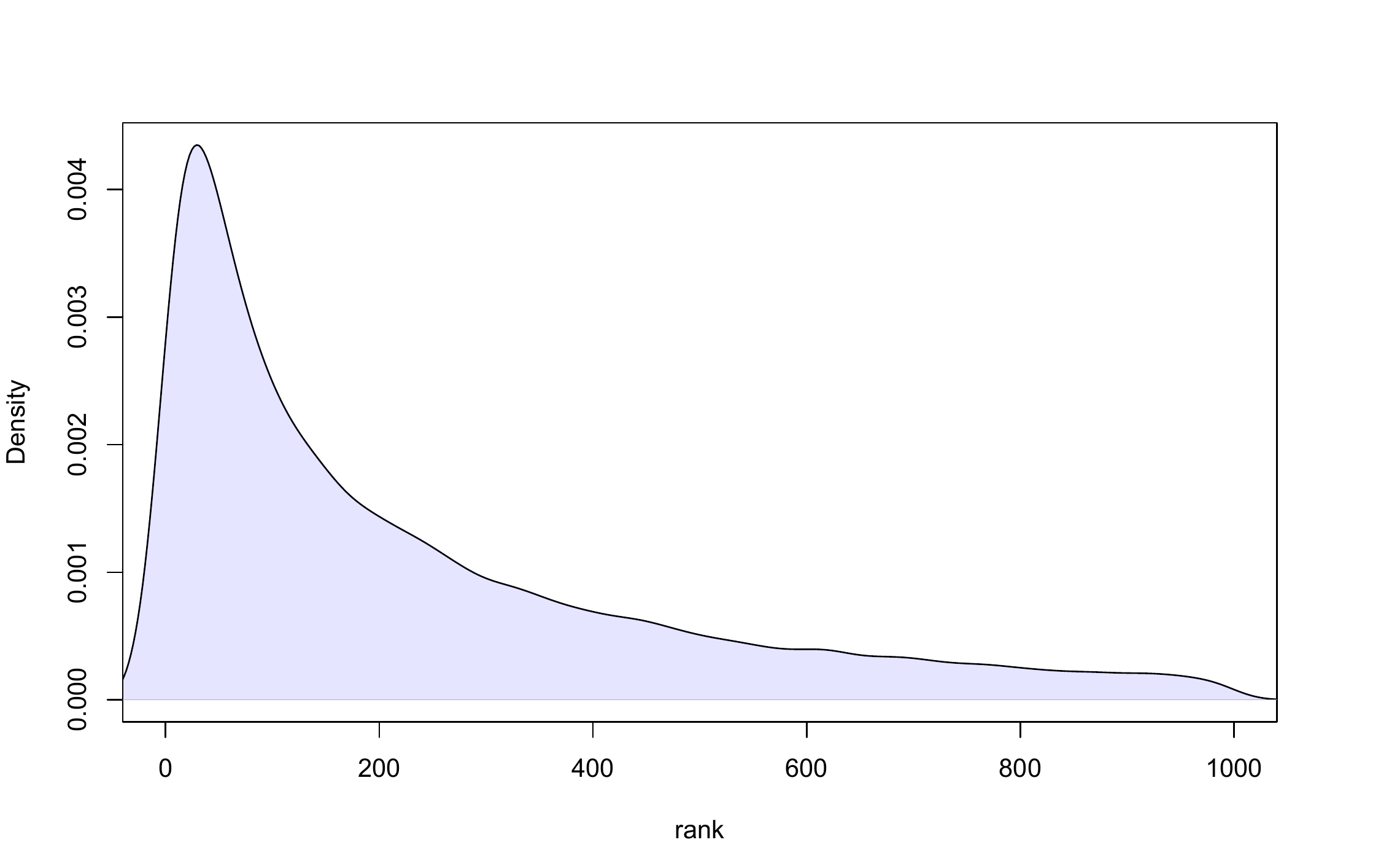}
\caption[]{Overall distribution of latent ranks for the name Innāya. The highest rank is 1; 1000 is the lowest.}
\label{imdi}
\end{centering}
\end{figure}

Figure \ref{imdi} shows the overall distribution for latent ranks associated with the name Innāya in all of the letters, as learned by our model.  Given the evidence that we have seen, our algorithm has learned that this name is generally associated with a very high-ranking individual (in this, and all other plots, the highest rank is 1, with 1000 being the lowest).  

\begin{figure}[htp]
\begin{centering}
\includegraphics[scale=.9]{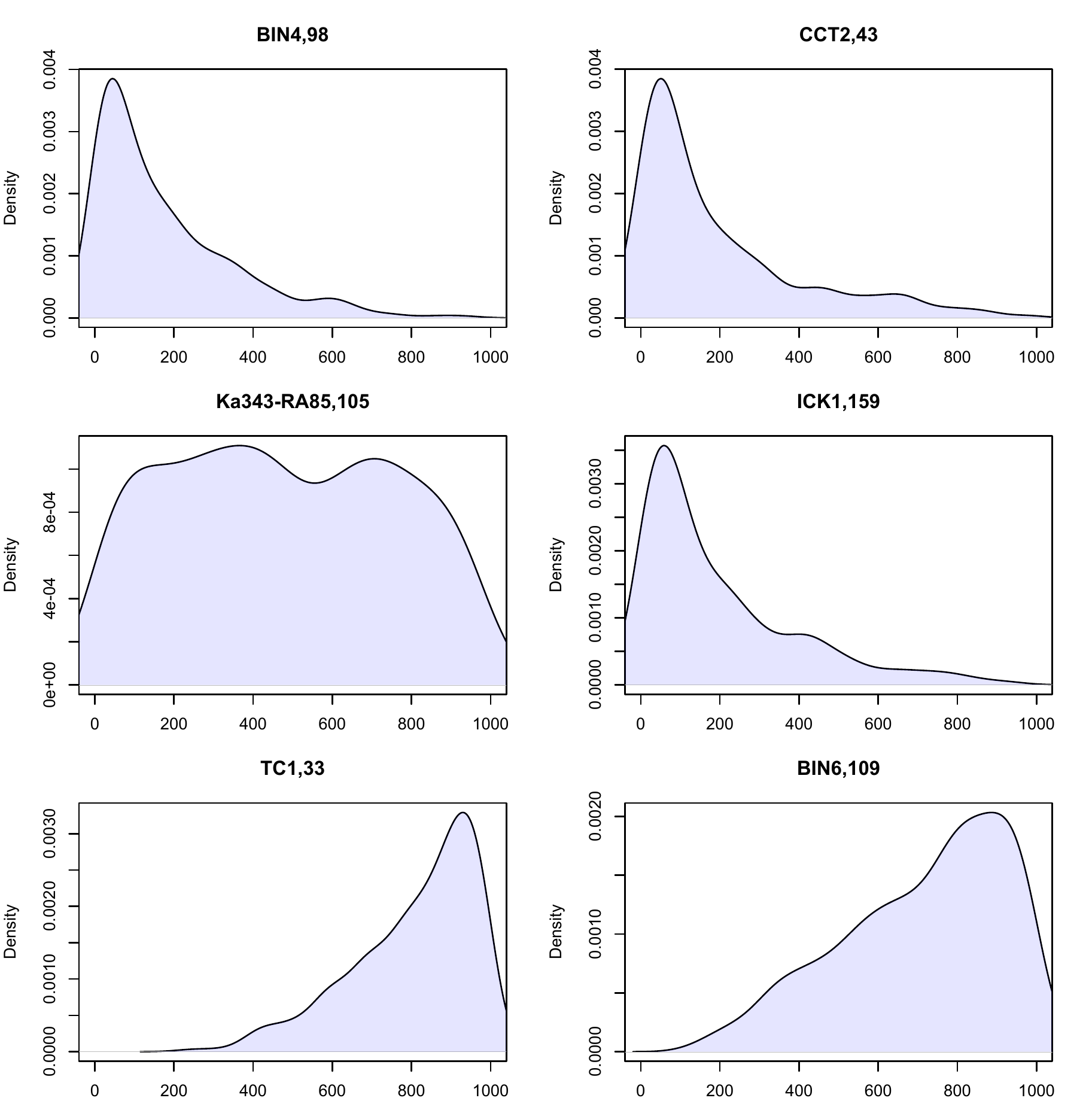}
\caption[]{Distribution over latent ranks associated with the name Innāya in various letters.  The highest rank is 1; 1000 is the lowest.}
\label{innmult}
\end{centering}
\end{figure}

If we look at the individual letters themselves, however, a more
nuanced picture emerges.  Figure \ref{innmult} plots the distribution
of latent entities for a set of 6 of the 190 letters in which Innāya
is mentioned.  While the majority of letters in this set recapitulate
the overall distribution shown above, several letters are noticeable
outliers.  For example, in letter TC1,33 and BIN6,109, our model has high belief that the real person associated with the name Innāya is not in fact the high-ranking individual at all, but rather a much lower ranked one.  Consulting these letters, we see that  Innāya is dominated by one or more individuals with a relatively low rank, which is not consistent with a high-ranking individual.  If we look at the intersection of the publicly available letters in our collection and those inspected by Michel (a total of 142 texts), we find that our  method agrees with Michel's assessment of the identity of Innāya in each specific letter 80.9\% of the time (discounting the level of agreement due to chance, this leads to a Cohen's $\kappa$ of 0.435).
These results clearly support the conclusions Michel has drawn for Innāya -- two individuals, at least, each with differing social networks and hierarchical ranks -- and provide evidence for the validity of our approach.

\section{Conclusion}

The case of Innāya illustrates one of the greatest obstacles facing the field of Old Assyrian studies to date. Before any definitive statements can be made about the organization and makeup of the Old Assyrian trade network on the basis of the texts found in Kültepe, we must first determine the scope and structure of the Old Assyrian colonial society.
Unfortunately, due to an active use of homonyms in the textual record, the scope of this colonial society has been obscured by a level of ambiguity too complex for any single specialist, or for that matter any group of specialists, to untangle.

As part of a solution, we present a method for aggregating small,
local pieces of information---pairwise social differences between
names in a cuneiform tablet---into a single underlying social order
that offers the best explanation for the data we have.  The latent
variable model that we design allows us to be clear about our
assumptions (the relationship between variables is encoded in the
structure of the model; whether or not one variable is allowed to
exert a direct effect on another is transparent).  These kinds of generative models also allow us to add other forms of evidence; one possible extension would tie the choice of latent entity for each name in a letter with all of the other text observed (so that, for example, if one Innāya is often associated with letters mentioning \emph{tin} while another trafficks in \emph{textiles}, we have further evidence that the two are different individuals).

 In applying this method to our Old Assyrian dataset, we are engaging in exploratory data analysis, offering informed hypotheses that are driven by data, and that our model believes are the most promising avenues for directed research by Assyriologists.  Grounding these hypotheses in the data allows us to return to the source of our induction -- the letters themselves -- and validate whether or not we have sufficient evidence to support our claims.  In our particular case, the agreement between our model's beliefs and those in the published literature are encouraging.  We leave to future work to explore the differences that remain.

\section{Acknowledgments}

We would like to thank Mogens Larsen, Thomas Hertel, and other contributors to the Old Assyrian Text Project for providing the digitized texts we analyze. The research reported in this article was supported in part by Google (through the Worldly Knowledge Project at CMU) and by an ARCS scholarship to D.B.

\bibliographystyle{plain}
\bibliography{../bibliography}

\begin{thebibliography}{1}

\bibitem{barjamovic_Ups_2012}
Gojko Barjamovic, Thomas Hertel, and Mogens~T. Larsen.
\newblock {\em Ups and Downs at Kanesh: Chronology, History and Society in the
  Old Assyrian Period}.
\newblock PIHANS (Publications de l'institut historique-arch{\'e}ologique
  n{\'e}erlandais de Stamboul), vol. 120. Nederlands Instituut voor het Nabije
  Oosten, Leiden, 2012.

\bibitem{Blei:2003:LDA:944919.944937}
David~M. Blei, Andrew~Y. Ng, and Michael~I. Jordan.
\newblock Latent {D}irichlet allocation.
\newblock {\em J. Mach. Learn. Res.}, 3:993--1022, March 2003.

\bibitem{michel_Innaya_1991a}
C{\'e}cile Michel.
\newblock {\em Inn{\=a}ya dans les Tablettes pal{\'e}o-assyriennes I: Analyse}.
\newblock Editions Recherche sur les Civilisations, Paris, {France}, 1991.

\bibitem{michel_Innaya_1991b}
C{\'e}cile Michel.
\newblock {\em Inn{\=a}ya dans les Tablettes pal{\'e}o-assyriennes II: Edition
  des textes}.
\newblock Editions Recherche sur les Civilisations, Paris, {France}, 1991.

\bibitem{michel_Alahum_2008}
C{\'e}cile Michel.
\newblock The {A}l{\=a}hum and {A}{\v s}{\v s}ur-takl{\=a}ku {A}rchives found
  in 1993 at {K}{\"u}ltepe {K}ani{\v s}.
\newblock {\em Altorientalische Forschungen}, 35:53--67, May 2008.

\bibitem{oatp}
{Old Assyrian Text Project}.
\newblock http://oatp.ku.dk.

\bibitem{veenhof}
K.~R Veenhof and J.~Eidem.
\newblock {\em Mesopotamia: The Old Assyrian Period}.
\newblock Orbis Biblicus Et Orientalis 160/5. Academic Press (Fribourg),
  Vandenhoeck and Ruprecht (G\"{o}ttingen), 2008.

\bibitem{wei90}
Greg C.~G. Wei and Martin~A. Tanner.
\newblock A {M}onte {C}arlo implementation of the {EM} algorithm and the poor
  man's data augmentation algorithms.
\newblock {\em Journal of the American Statistical Association}, 1990.

\end{thebibliography}

\end{document}